\documentclass[12pt]{article}

\pdfoutput=1

\usepackage{graphicx}
\usepackage{times}
\usepackage{comment}

\usepackage{tabularray}
\SetTblrInner{colsep=4.5pt,rowsep=1pt}
\usepackage{xcolor}

\title{Valley-controlled photoswitching \\ of metal-insulator nanotextures}

\author
{Hannes Böckmann,$^{1,2\ast}$ Jan Gerrit Horstmann,$^{1,2}$ Felix Kurtz,$^{1,2}$\\
Manuel Buriks,$^{3}$ Karun Gadge,$^{3}$ Salvatore R. Manmana,$^{3}$\\
Stefan Wippermann,$^{4}$ Claus Ropers$^{1,2}$ \\
\\
\normalsize{$^{1}$Max Planck Institute for Multidisciplinary Sciences, Göttingen, Germany}\\
\normalsize{$^{2}$4th Phys. Inst., Solids and Nanostructures, University of Göttingen, Germany}\\
\normalsize{$^{3}$Institute for Theoretical Physics, University of Göttingen, Germany}\\
\normalsize{$^{4}$Faculty of Physics, Philipps University of Marburg, Germany}\\
\\
}

\date{}

\begin{document} 
\setlength{\emergencystretch}{20pt}

\baselineskip24pt


\maketitle 
\section*{Abstract}

Spatial heterogeneity and phase competition are hallmarks of strongly-correlated materials \cite{dagotto_nanoscale_2003,ahn_strain-induced_2004}, directly connected to intriguing phenomena such as colossal magnetoresistance \cite{uehara_percolative_1999,fath_spatially_1999,lai_mesoscopic_2010} and high-temperature superconductivity \cite{sharma_percolative_2002,park_electronic_2009}. Active control over phase textures further promises tunable functionality on the nanoscale \cite{cao_strain_2009}. Light-induced switching of a correlated insulator to a metallic state is well established. However, optical excitation generally lacks the specificity to select sub-wavelength domains and control final textures. Here, we employ valley-selective photodoping to drive the domain-specific quench of a textured Peierls insulator. Polarized excitation leverages the anisotropy of quasi-one-dimensional states at the correlated gap to initiate an insulator-to-metal transition with minimal electronic heating. We find that averting dissipation facilitates domain-specific carrier confinement, control over nanotextured phases, and a prolonged lifetime of the metastable metallic state. Augmenting existing manipulation schemes, valley-selective photoexcitation will enable the activation of electronic phase separation beyond thermodynamic limitations, facilitating optically-controlled hidden states, engineered heterostructures, and polarization-sensitive percolation networks.

\section*{Main}

\begin{figure}[t]%
\centering
\includegraphics[width=1\linewidth]{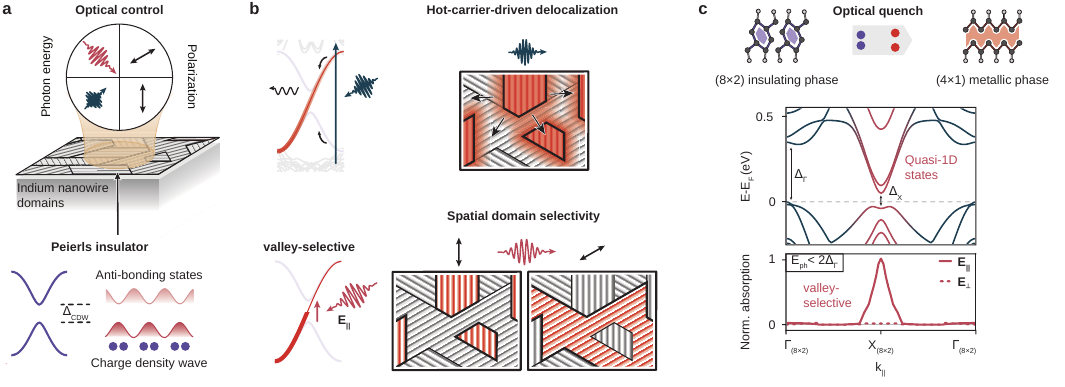}
\caption{\textbf{Optical surface electronic texture control via valley-selective photodoping.} (\textbf{a}), Top, schematic of optical control parameters that drive selective surface domain switching. Bottom, the Peierls transition in a one-dimensional atomic chain yields the formation of a periodic lattice distortion and charge density wave (CDW) with a characteristic energy gap ($\Delta_{CDW}$). (\textbf{b}), Top, light-induced electronic quench of the CDW phase. A population of states at the CDW gap, mediated by higher-lying surface bands, is accompanied by electronic and lattice heating and yields a homogeneous phase change across the surface via delocalized energetic charge carriers. Bottom, direct band gap excitation (valley-selective) minimizes dissipation which manifests in domain-specific switching (red wires) with pronounced phase coexistence and local carrier confinement. (\textbf{c}), Top, Optical transition between the insulating (8$\times$2) and metallic (4$\times$1) phase of indium nanowires. Center, density-functional-theory (DFT) calculated electronic band structure of the Si(111)(8$\times$2)-
In phase (see Methods for details).  $\Delta_{\Gamma}$ and $\Delta_{X}$ denote CDW gaps, leading to the formation of energetic valleys in the band structure. Bottom, polarized optical absorption. X-valley selectivity at parallel
polarization for low excitation energies enables orientation-specific switching. $k_{\vert\vert}$ is the momentum vector along Brillouin zone high-symmetry points with respect to the wire direction. }
\label{fig1}
\end{figure}

The interactions of electronic, orbital, spin and nuclear degrees of freedom govern the emergence of symmetry-broken functional states in solids. Light allows for tilting the balance between distinct correlated states and phases \cite{kogar_light-induced_2020} and enables the control over final states by precisely tuning the optical interaction (Fig. \ref{fig1}a, top). A prototypical scenario is given by the optical quench of an electronic density modulation and lattice distortion in charge-density-wave (CDW) materials \cite{sciaini_electronic_2009,morrison_photoinduced_2014,eichberger_snapshots_2010,danz_ultrafast_2021}.
In the prevalent case of a Peierls insulator \cite{peierls_quantum_2001}, the interplay of quasi-one-dimensional (1D) electronic states at the Fermi energy and the lattice instability lead to the opening of a band gap ($\Delta_{CDW}$) and the formation of electronic valleys in a metal-to-insulator transition (Fig. \ref{fig1}a, bottom) \cite{imada_metal-insulator_1998}. Photodoping of occupied bonding and unoccupied antibonding states at the band edges, in turn, collapses the band gap and transiently reverses the phase change \cite{frigge_optically_2017,bockmann_mode-selective_2022}.

Whereas such optical switching frequently occurs on femtosecond timescales, it is largely indiscriminate with respect to the induced electronic transitions. In particular, correlated states near the gap are generally populated indirectly via rapid relaxation from optically accessible higher-lying bands (Fig. \ref{fig1}b, top) \cite{sciaini_electronic_2009,nicholson_beyond_2018}. The deposited energy per lifted carrier, therefore, far exceeds that required for a minimally invasive phase transformation, which leaves the system with substantial lattice and electronic heat. A valley-polarized population from resonant gap excitation \cite{mak_control_2012}, on the other hand, may largely prevent dissipation to promote a minimum-energy transition pathway (Fig. \ref{fig1}b, bottom). Moreover, close to the photon-energy threshold, the phase change is expected to become particularly susceptible to local variations of the real-space microstructure and domain distributions \cite{morrison_photoinduced_2014, johnson_ultrafast_2022}.\\
\indent Realizing such conditions, in this work, we employ valley-selective photodoping of correlated electronic states to demonstrate control over the domains and texture of a quasi-one-dimensional Peierls insulator. Specifically, we exploit the anisotropic absorption of nanowire domains by tuning both the photon energy and the polarization to the transition matrix elements most strongly coupled to the structural transformation. The reduction in electronic excess energy facilitates a minimum-energy pathway to domain-specific switching and a prolonged metastable lifetime. This selection of the carrier energy allows for a spatial control over the phase transition, from delocalized carriers and homogeneous switching at high photon energy to locally-confined carriers and domain-specific switching for resonant gap excitation. The here demonstrated photoinduced formation of metallic surrounded by insulating nano-domains represents the controlled preparation of nanoscale electronic textures, suggesting avenues for optically engineered electronic percolation and quantum-confinement in 'Peierls heterostructures'.

\section*{Valley-specific excitation of Peierls-distorted atomic wires}

Atomic wires formed by indium atoms on the (111) face of silicon \cite{yeom_instability_1999} are a prominent model system with near-ideal quasi-1D electronic properties. The corresponding surface reconstruction in a metallic (4$\times$1) room-temperature phase is characterized by a parallel arrangement of atomic zigzag chains (Fig. \ref{fig1}c, top) \cite{lander_surface_1965}. Upon cooling below the critical temperature $T_c=125\,K$, a triple-band Peierls transition \cite{ahn_mechanism_2004} transforms the system into the insulating (8$\times$2) hexagon phase. A doubling of lattice periodicity is brought about by shear and rotary distortions, leading to the formation of interchain and intrachain covalent bonds within the coupled indium chains, respectively \cite{wippermann_entropy_2010}. The phase change is equally reflected in the electronic band structure by the opening of band gaps and concomitant formation of energetic valleys  (Fig. \ref{fig1}c, center). Lateral shearing of zigzag chains causes a band gap opening at the surface Brillouin zone $\Gamma$-point ($\Delta_{\Gamma}$) \cite{gonzalez_metalinsulator_2005}, while rotation-induced dimerization of outer indium atoms yields a band gap at the X-point ($\Delta_{X}$) \cite{wippermann_entropy_2010}.


Notably, the formation of a supercooled metallic (4$\times$1) phase can be triggered by light without reaching the phase transition temperature. In this process, optically generated electrons and holes rapidly scatter towards correlated electronic states, quenching the insulating band gap and lifting the structural distortion on a $350\,\mathrm{fs}$ timescale \cite{frigge_optically_2017}. While the ultrafast transition involves a directed coherent nuclear motion driven by valley-specific couplings between carrier populations and well-defined unit cell distortions \cite{wippermann_entropy_2010,jeckelmann_grand_2016,bockmann_mode-selective_2022}, the associated band structure dynamics were thus far found to be independent of the incident photon energy \cite{nicholson_beyond_2018,chavez-cervantes_charge_2019}. However, the inherent anisotropy of bands at the insulating gap holds the potential for spatially distinct final phase textures, which have not been explored \cite{chandola_structure_2009}.\\
In Fig. \ref{fig1}c, bottom, we display the normalized optical absorption of indium nanowires. For low photon energies (less than twice the $\Gamma$-point gap), we find that absorption is largely limited to transitions near the X-valley (at the insulating gap $\Delta_X$) and to a polarization parallel to the nanowire direction (absorption at higher energies is shown in Extended data Fig. \ref{abs_highE}). In this regime, a 'valley-selective' transition from bonding to anti-bonding states enables orientation-dependent optical switching (see also Extended data Fig. \ref{band_structure}; Supplementary information for tight-binding calculations of momentum-resolved absorption).

\section*{Creating coexistent electronic phases by polarization-selective switching}

We test this notion experimentally by using time-resolved low-energy electron diffraction (ULEED) combined with wavelength- and polarization-controlled excitation (see Methods). In short, ULEED employs backscattering diffraction of electron pulses from surfaces to probe optically-induced changes of the atomic-scale structure \cite{gulde_ultrafast_2014,vogelgesang_phase_2018,storeck_structural_2020,horstmann_coherent_2020}. In the present system, we monitor the phase transformation via changes in the intensity of characteristic diffraction peaks. 
Specifically, at a base temperature of $T=60\,\mathrm{K}$, indium nanowires in the (8$\times$2) CDW phase undergo a pump-induced ultrafast transition into the metastable (4$\times$1) structure, which is subsequently probed by a photoemitted electron pulse. After the pump pulse, we observe an intensity decrease and increase of (8$\times$2) and (4$\times$1) diffraction features, respectively.


\begin{figure}[t]%
\centering
\includegraphics[width=1\textwidth]{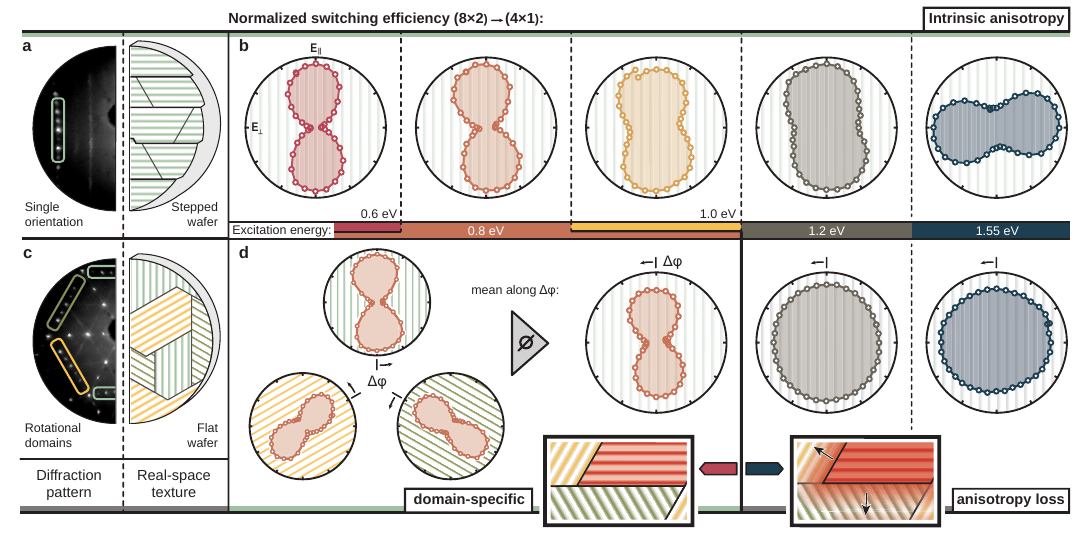}
\caption{
\textbf{Polarization, photon energy and domain texture dependency of the (8$\times$2) to (4$\times$1) switching efficiency.} (\textbf{a}), LEED image and real-space sketch of parallel-oriented indium atomic wires on a stepped Si wafer surface with a $2^\circ$-miscut relative the (111) plane. (\textbf{b}), Polarization-dependent switching efficiency for increasing photon energy, at maximum diffraction spot suppression ($1-I(8\times2)_{\Delta t=40\,\mathrm{ps}}/I(8\times2)_{\Delta t=-150\,\mathrm{ps}}$). The in-plane electric field component is depicted relative to the nanowire direction. Incident fluences from left to right: 1.04, 0.59, 0.62, 1.09, 1.40$\,$mJ$\,$cm$^{-2}$. (\textbf{c}), Three-fold symmetric rotational domains on a flat Si(111) wafer surface. The LEED image comprises a superposition of domain-specific reflexes.
(\textbf{d}), Polarization-dependent switching of rotational domains. $\Delta\varphi$ denotes the angle between in-plane electric field and the nanowire orientation. Incident fluences from left to right: 1.60, 1.40, 2.03$\,$mJ$\,$cm$^{-2}$. The transition between valley-selective to multi-band excitation manifests in a loss of the intrinsic anisotropy. }
\label{fig2}
\end{figure}

As a central finding, we observe that the switching efficiency of the atomic wires is highly dependent on the incident electric-field polarization.
We systematically trace the photon-energy-resolved intensity suppression of (8$\times$2) diffraction features at a fixed time delay ($\Delta t=40\,$ps) as a function of polarization (analyzed diffraction features are sketched in Extended data Fig. \ref{Ext_DiffrFeat}). For this purpose, a preferential orientation of the nanowires on the Si(111) surface is imposed by a linear step gradient on a $2^\circ$-miscut wafer (Fig. \ref{fig2}a). At infrared (IR) driving wavelengths (Fig. \ref{fig2}b, left), the observed anisotropy aligns with the nanowire orientation, which specifically points towards the direct absorption by quasi-1D electronic states close to the $X$ point in the surface Brillouin zone (see Fig. \ref{fig1}c, bottom). This assignment is corroborated by the fingerprint of the involved electronic states, given by their preferential coupling to the vibrational rotation mode that modulates the structural distortion upon excitation \cite{wippermann_entropy_2010,jeckelmann_grand_2016,bockmann_mode-selective_2022}. Indeed, we observe an increased amplitude of the rotational mode in complementary coherent control experiments at IR wavelengths (see Supplementary information).\\
\indent In the near IR, the anisotropy gradually reverses towards perpendicular polarization (Fig. \ref{fig2}b, right), where the population of correlated states is largely mediated by higher-energy surface bands, spread across the entire surface Brillouin zone. Our findings are qualitatively consistent with linear reflectance anisotropy \cite{chandola_structure_2009} and IR spectroscopy \cite{chung_optical_2010} studies. Yet, the non-linear characteristic of the phase transition efficiency with absorbed fluence further enhances the differences in absorption \cite{horstmann_coherent_2020}. \\
Harnessing the measured intrinsic anisotropy, we next demonstrate real-space texture control in the rotational domain structure formed naturally on a non-stepped Si(111) surface (Fig. \ref{fig2}c) (domain size range: $10^3-10^4\,\mathrm{nm}^2$ \cite{shim_true_2023}). Here, the nanowire orientation is directly linked to the diffraction reflex angle in the LEED image, providing orientation-resolved access to the phase change within the domain texture. In order to illustrate the dependence of the switching efficiency with the in-plane electric field component, the domain-specific anisotropy is depicted with respect to the individual nanowire orientation (Fig. \ref{fig2}d, left). For IR excitation, the polarization dependence agrees with the intrinsic anisotropy, i.e., with that measured on a stepped surface, which demonstrates domain-specific absorption and switching. In stark contrast, at shorter near-IR wavelengths, we find a surprisingly isotropic response (Fig. \ref{fig2}d, right), strongly deviating from the intrinsic anisotropy. From these data, we conclude that the homogeneous global response of all domains stems from the delocalizing transfer of energetic photoexcited charge carriers across domain boundaries, effectively eliminating the intrinsic domain response. The strong dependence on photon energy is further attributed to the increasing amount of optical excess energy in the electron system after the insulator-to-metal transition, allowing for a higher mobility across potential barriers between metallic and neighboring insulating domains.

\section*{Enhanced metastability from minimized thermal fluctuations}

\begin{figure}
\centering
\includegraphics[width=0.6\textwidth]{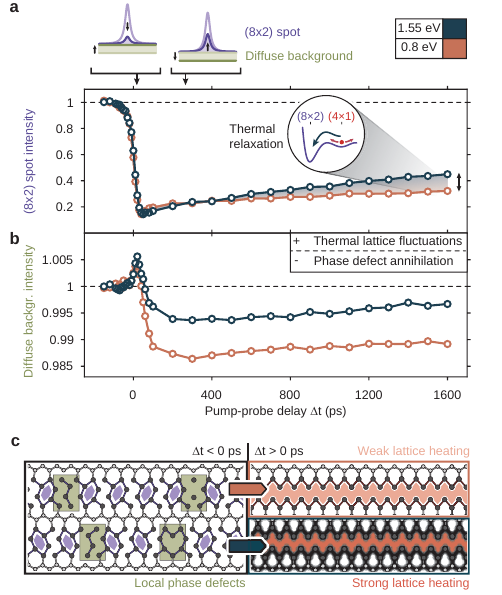}
\caption{\textbf{Thermal relaxation of the metastable (4$\times$1) phase, driven by optical excess energy.} (\textbf{a}), Integrated (8$\times$2) diffraction spot intensity as function of pump-probe delay $\Delta t$ at 0.8 eV and 1.55 eV incident photon energy on a stepped wafer. The light polarization was chosen parallel and perpendicular to the nanowire orientation, respectively, corresponding to the maximum switching efficiency at fluences $F_{0.8}$=1.95$\,$mJ$\,$cm$^{-2}$ and $F_{1.55}$=3.12$\,$mJ$\,$cm$^{-2}$. For identical intensity suppression, the relaxation into the (8$\times$2) ground state accelerates with photon energy due to increasing thermal lattice fluctuations, following the phase transition (see inset). (\textbf{b}), Pump-induced dynamic disorder increases the diffuse background intensity, counteracted by a reduction in static disorder from enhanced phase homogeneity. (\textbf{c}), Left, real-space sketch of indium atomic wires in the (8$\times$2) phase, showing characteristic phase defects (alongside phase boundaries \cite{song_dynamical_2019}) which cause diffuse background scattering. Right, Photoinduced (4$\times$1) structure exhibits an increased phase homogeneity. Thermal lattice fluctuations cause an accelerated relaxation for increasing photon energy.}
\label{fig3}
\end{figure}
Beyond charge carrier confinement, we examine the implications of lattice heating from photon excess energy on the metastable state. Specifically, we track the temporal evolution of the (8$\times$2) spot intensity at identical initial suppression on a stepped wafer and observe a prolonged lifetime from near-IR to IR excitation (Fig. \ref{fig3}a). We ascribe this difference to photon-energy-dependent thermal fluctuations, driving the relaxation via over-the-barrier transitions (Fig. \ref{fig3}a, inset) \cite{tao_nature_2016}. Interestingly, we find that the reduction in lattice heating yields an increasingly ordered metastable (4$\times$1) phase, in which (8$\times$2) phase defects are annihilated \cite{zhang_atomic_2011,kim_topological_2012}.\\
The diffuse scattering background reflects this through an unusual pump-induced intensity reduction, becoming more pronounced at low photon energy (Fig. \ref{fig3}b). While thermal fluctuations lead to a transient increase of elastic scattering, the annihilation of phase defects decreases it, as the surface is driven into a more homogeneous phase texture.
Hence, our findings imply that valley-selective excitation approaches a pathway with minimized entropy for the phase transition and leaves the system in a highly-ordered state from which thermal relaxation is suppressed (Fig. \ref{fig3}c).

\section*{Discussion and Outlook}

\begin{figure}[]
\centering
\includegraphics[width=1\textwidth]{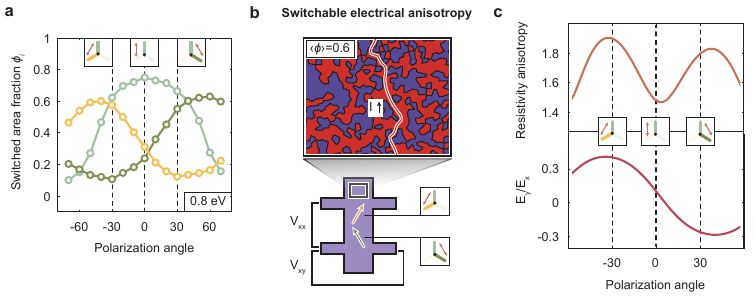}
\caption{\textbf{Predicted functionality of optically engineered Peierls heterostructures.} (\textbf{a}), Measured rotational domain-resolved surface area fraction, switched into the (4$\times$1) phase at 0.8 eV photon energy (see also Fig. \ref{fig2}d). The insets indicate particularly domain-selective optical polarization states. (\textbf{b}), Electronic transport in a percolated domain texture within a Hall bar geometry. Red, metallic (4$\times$1); blue, insulating (8$\times$2) domains. Selective domain switching could be used to control the transversal anisotropic charge flow. (\textbf{c}), Top, calculated anisotropy along the conducting network principal axes with incident polarization, using experimental data. Bottom, transversal to longitudinal voltage ratio across the Hall bar. The voltage polarity can be optically switched by deselecting individual wire orientations. }
\label{fig4}
\end{figure}

From these observations, an intuitive picture of the spatio-temporal phase change dynamics in the regime of valley-selective photodoping emerges. IR photon absorption effectively confines charge carriers to rotational nanodomains, selected by the incident polarization (Fig. \ref{fig4}a). The population and depopulation of strongly correlated electronic states at the X-valley collapses the band gap ($\Delta_X$ in Fig. \ref{fig1}c) in an insulator-to-metal transition within about 200 fs \cite{nicholson_beyond_2018}, localizing the low-energy photoexcited charges to the absorbing domain. As thermal fluctuations are suppressed, the metastability likely persists far beyond the commonly observed few-nanosecond lifetime \cite{hafke_condensation_2019,horstmann_coherent_2020}.

We believe that optically engineered phase textures expand the paradigm of light-induced phenomena and enable a polarization-sensitive control of functional states in real space. Therefore, motivated by our experimental findings, we theoretically explore a tangible implementation of 'Peierls heterostructures'. Polarization-specific switching can yield optically-controlled electronic properties in correlated percolation networks. While homogeneous switching results in a surface texture with isotropic conductivity, i.e., an equal fraction of metallic nanowire orientations, domain-specific switching promises tunable anisotropy when the incident polarization and fluence is adjusted accordingly (further details are given in Methods). Specifically, we propose the optical steering of current flow, as manifested by the build-up of a transversal voltage in a Hall bar geometry (Fig. \ref{fig4}b). The characteristics of this measurement are modeled, based on experimental data for the orientational composition of nanowires in a metallic percolation network and the intrinsic conductivity of indium atomic chains \cite{kanagawa_anisotropy_2003}. We find that electrical anisotropy along the principal axes is peaked when one orientation is deselected by choosing a polarization perpendicular to the wire direction (Fig. \ref{fig4}c, top). In this scenario, the preferential transport along the wire direction yields a lateral deflection of charge carriers, resulting in a transversal voltage with tunable polarity (Fig. \ref{fig4}c, bottom). 
Apart from the shown example, a possible implementation of optically tailored electronic properties is given by the transient formation of isolated metallic grains at parallel polarization, resulting in the quantization of electronic states within the insulating band gap (see Extended data Fig. \ref{LDOS} for tight-binding simulations of the resulting local density of states).\\
\indent In conclusion, our results demonstrate the quench of a Peierls insulator by means of valley-selective optical transitions that specifically address strongly-correlated orbitals. In this way, we achieve deep sub-wavelength precision in an optically-driven phase change and gain control over the real-space texture. Our results suggest an important role of optical excess energy in photo-induced phenomena with immediate ramifications for the ability to prepare and manipulate coexisting electronic phases. 
The inherent nature of Peierls physics at the core of this quasi-1D system implies a broader relevance to materials, wherein changes between electronic phases are instigated by distinct low-symmetry electronic states. As such, we believe that polarization-sensitive switching offers a versatile framework for patterning domain structures and controlling nano-engineered functionalities across various exotic electronic systems, such as correlated oxides \cite{morrison_photoinduced_2014,fausti_light-induced_2011,tao_nature_2016,wegkamp_instantaneous_2014,ronchi_nanoscale_2022,mcleod_nanotextured_2017,johnson_ultrafast_2022}, Weyl semimetals \cite{sie_ultrafast_2019,guan_manipulating_2021}, and materials exhibiting electronic nematicity \cite{fernandes_intertwined_2019,nie_charge-density-wave-driven_2022, li_rotation_2022,zhao_cascade_2021}. Hence, optical preparation of electronic phase separation introduces an additional degree of freedom in light-induced switching.

\newpage

\section*{Methods}\label{Methods}

\subsection*{Density functional theory calculations}\label{DFT}

We performed DFT simulations within the local-density approximation (LDA) \cite{ceperley1980} as implemented within the Vienna \textit{ab-initio} simulation package (VASP) \cite{kresse1996}. The electronic structure is described by projector-augmented wave potentials \cite{bloechl1994} with a plane wave basis set limited to a cutoff energy of 250$\,$eV. The surface was modeled using periodic boundary conditions and a slab with three bilayers of silicon. Si dangling bonds at the bottom layer were saturated with hydrogen. To determine the ground-state electronic structure, a $2\times8\times1$ Monkhorst-Pack mesh was used to sample the Brillouin zone of the Si(111)-(8 $\times$ 2)In structure, corresponding to 256 k-points in the Si(111)-($1 \times 1$) surface unit cell. The band structure calculations were performed using 40 (10) k-points along the $\Gamma-X$ ($X-M$) directions, respectively.\\
The orbital character of the bonding and anti-bonding states, as illustrated in Extended data Fig. \ref{band_structure}, was determined by integrating the electronic density of states along the $X - M$ direction and over the topmost 4 valence bands and lowest 4 conduction bands, respectively. The resulting densities were plotted in Extended data Fig. \ref{band_structure} at an isovalue of 0.033 $e^-\mathring{\mathrm{A}}^{-3}$.\\
Consistent with the numerical approach outlined in Ref. \cite{chandola_structure_2009}, the oscillator strength was calculated in independent particle approximation (IPA) from the squared transition matrix elements ($\|M_{if}\|^2$) between all occupied (i) and unoccupied (f) surface bands within the Brillouin zone along the depicted high-symmetry points at parallel and perpendicular polarization with respect to the wire direction. To this end, we considered 240 occupied and 120 empty bands. The calculated values at all momenta are energy integrated and normalized to the overall maximum. 

\subsection*{Ultrafast LEED and optical setup}\label{ULEED}

\begin{figure}[h]%
\centering
\includegraphics[width=0.5\textwidth]{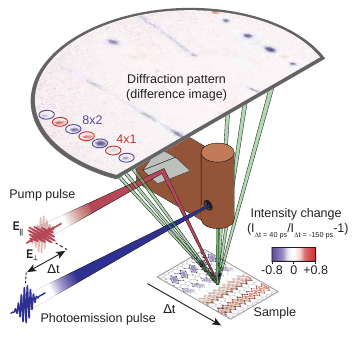}
\caption{\textbf{Experimental setup of ultrafast LEED.} Ultrashort electron pulses, photoemitted from an electron gun, probe the microscopic surface structure after pump-probe delay $\Delta t$. The optical quench of the structural distortion manifests in a loss/gain of (8$\times$2)/(4$\times$1) reflex intensity.}
\label{ULEED_sketch}
\end{figure}

Ultrafast low-energy electron diffraction (ULEED) is a technique for the investigation of structural dynamics at solid-state surfaces by means of an optical-pump/electron-probe scheme (Fig. \ref{ULEED_sketch}) \cite{gulde_ultrafast_2014,vogelgesang_phase_2018,storeck_structural_2020,horstmann_coherent_2020}. It combines the high surface sensitivity of photoemitted low-energy electrons in backscattering geometry with ultrafast optical excitation to follow the evolution of non-equilibrium surface structures .\\
The demonstrated high temporal and momentum resolution is achieved with a home-built laser-driven electron gun consisting of a nanometric tungsten tip as well as four metal electrodes (outer diameter of 2$\,$mm). Ultrashort electron pulses are generated from localized two-photon photoemission by illuminating the tip apex with femtosecond laser pulses (central wavelength $\lambda_c=400\,$nm, pulse duration $\tau_p = 45\,$fs, pulse energy $E_p = 30\,$nJ) at a repetition rate of 100$\,$kHz. The small diameter of the electron gun allows for a small sample distance without blocking the backscattered electrons, resulting in pulse durations down to 16$\,$ps. Detected electrons are amplified and recorded by a combination of a chevron microchannel plate, a phosphor screen, and a scientific complementary metal–oxide–semiconductor camera.\\
In the described experiments, the surface is excited with a wavelength-tunable optical pump pulse from an optical parametric amplifier ($\lambda_c$=800, 1240, 1550, 2066$\,$nm, $\hbar\omega$=1.55, 1.0, 0.8, 0.6$\,$eV, $\Delta\tau$ = 232$\,$fs) or with a fixed pulse wavelength from a Yb:YAG amplifier system ($\lambda_c=1030\,\textrm{nm}$, $\hbar\omega$=1.2$\,$eV, $\Delta\tau$=212$\,$fs). The surface is probed by a photoemitted 80$\,$eV electron pulse with a beam diameter of ($\approx$ 80$\times$80$\,\mu$m$^2$ (FWHM)).

\subsection*{Sample preparation}\label{sample_prep}

All experiments in this work have been conducted under ultra-high-vacuum conditions (base pressure $p<2\times10^{-10}$\,mbar), thus minimizing the effect of adsorbate-related (8$\times$2) ground state recrystallization from the supercooled (4$\times$1) phase \cite{wall_atomistic_2012}. The silicon substrate (phosphorus-doped wafers with resistivity R=0.6–2$\,\Omega$ cm), showing only a single wire orientation, was miscut by $2\,^{\circ}$ towards the [-1 -1 2] direction to create a high surface step density. The oriented steps effectively confine the atomic indium wires along one crystallographic direction. On the other hand, wafers cut along the (111) face of silicon exhibit much larger terraces, where all three crystallographically equivalent orientations are equally found. All samples were cleaned by flash-annealing at 1350$\,^{\circ}C$ for five seconds through direct current heating. Subsequently, 1.2 monolayers of indium were deposited onto the resulting Si(111) (7$\times$7) surface reconstruction at room temperature and annealed at T=400$\,^{\circ}C$ for 300 s. The resulting Si(111) (4$\times$1)-In phase was inspected in our ultrafast LEED setup and subsequently cooled to a base temperature of T=60$\,$K using an integrated continuous-flow helium cryostat. The phase transition between the high-temperature (4$\times$1) and low-temperature (8$\times$2) phases was observed at a temperature of 125$\,$K.

\subsection*{Tight-binding simulations}\label{tb_sim}

\subsubsection*{Model for homogeneous systems}

\begin{figure}[!h]
  \centering
    \includegraphics[width=0.4\columnwidth]{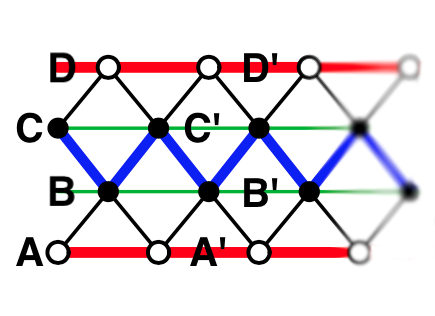}
\caption{\textbf{Lattice site indexing for two unit cells in the symmetric (high-temperature) configuration.} [modified from Ref~\cite{jeckelmann_grand_2016}].}
\label{fig:lattice_index}
\end{figure}

Following Ref.~\cite{jeckelmann_grand_2016} we choose a quasi-1D tight-binding Hamiltonian with four coupled parallel chains. 
Indexing of the lattice sites for two $(4\times1)$ unit cells is shown in Fig.~\ref{fig:lattice_index}. 
Due to the symmetry in this setup we denote this as the symmetric configuration or metallic phase (see the discussion further below on the band structure of the system). 
Note that this configuration corresponds to the high temperature phase.
We will add a distortion to the lattice, which leads to a metal-insulator transition and doubled lattice periodicity, similar to a Peierls transition~\cite{peierls_quantum_2001}. 
This corresponds to the $(8\times2)$ low-temperature configuration.

We discuss our approach on the simpler symmetric configuration, which consists of 4 sites per unit cell.  
Let $m \in \{1,...,N\}$ be the index for the unit cell and $\alpha \in \{A,B,C,D\}$ the intracell index. 
A general tight-binding Hamiltonian in real space is denoted as:
\begin{equation}
    H = \sum_{(m,\alpha),\sigma} \epsilon^{\phantom{\dagger}}_{(m,\alpha)} c^{\dagger}_{(m,\alpha),\sigma} c^{\phantom{\dagger}}_{(m,\alpha),\sigma}
    - \sum_{<i,j>,\sigma} t^{\phantom{\dagger}}_{ij} \left( c^{\dagger}_{i,\sigma} c^{\phantom{\dagger}}_{j,\sigma} + c^{\dagger}_{j,\sigma} c^{\phantom{\dagger}}_{i,\sigma}\right) \, ,
    \label{eq:ham_general}
\end{equation}
where $\sigma$ is the spin index, $\epsilon_{(m,\alpha)}$ is the on-site energy on lattice site $(m,\alpha)$, and $t_{ij}$ is the hopping strength between lattice sites $i$ and $j$, where we simplify the notation by setting $(m,\alpha)=:i$ and $(m',\alpha')=:j$ in the second sum.
$c_{i,\sigma}^{\dagger}$ is the usual fermionic annihilation (creation) operator for an electron with spin $\sigma$ on lattice site $i$.
We omit the spin index of the two spin directions in the following, as they are treated independently. In position space, this leads to the following matrix representation of the Hamiltonian:

\begin{equation}
\label{eq:ham_position}
 H_{\mathrm{reduced}} =  
  \left(\begin{tblr}{
    colspec = {cccc|[2pt]cccc},
    cell{1}{2} = {gray7},
    cell{2}{1} = {gray7},
    cell{2}{5} = {gray7},
    cell{5}{2} = {gray7},
    cell{1}{5} = {red8},
    cell{5}{1} = {red8},
    cell{2}{3} = {blue8},
    cell{3}{2} = {blue8},
    cell{2}{6} = {green8},
    cell{6}{2} = {green8},
    cell{2}{7} = {blue8},
    cell{7}{2} = {blue8},
    cell{3}{7} = {green8},
    cell{7}{3} = {green8},
    cell{3}{4} = {gray7},
    cell{4}{3} = {gray7},
    cell{4}{8} = {red8},
    cell{8}{4} = {red8},
    cell{4}{7} = {gray7},
    cell{7}{4} = {gray7},
    cell{5}{6} = {gray7},
    cell{6}{5} = {gray7},
    cell{6}{7} = {blue8},
    cell{7}{6} = {blue8},
    cell{7}{8} = {gray7},
    cell{8}{7} = {gray7},
    cell{1}{1} = {yellow9},
    cell{2}{2} = {brown9},
    cell{3}{3} = {brown9},
    cell{4}{4} = {yellow9},
    cell{5}{5} = {yellow9},
    cell{6}{6} = {brown9},
    cell{7}{7} = {brown9},
    cell{8}{8} = {yellow9},
  }
     \epsilon_{AA} & t_{AB} & t_{AC} & t_{AD} & t_{AA'} & t_{AB'} & t_{AC'} & t_{AD'} \\
     t_{BA} & \epsilon_{BB} & t_{BC} & t_{BD} & t_{BA'} & t_{BB'} & t_{BC'} & t_{BD'} \\
     t_{CA} &  t_{CB} & \epsilon_{CC} & t_{CD} & t_{CA'} &  t_{CB'} & t_{CC'} & t_{CD'} \\
     t_{DA} &  t_{DB} &  t_{DC} & \epsilon_{DD} & t_{DA'} &  t_{DB'} &  t_{DC'} & t_{DD'} \\
     \hline[2pt]
     t_{A'A} & t_{A'B} & t_{A'C} & t_{A'D} & \epsilon_{A'A'} & t_{A'B'} & t_{A'C'} & t_{A'D'} \\
     t_{B'A} & t_{B'B} & t_{B'C} & t_{B'D} & t_{B'A'} & \epsilon_{B'B'} & t_{B'C'} & t_{B'D'} \\
     t_{C'A} &  t_{C'B} & t_{C'C} & t_{C'D} & t_{C'A'} &  t_{C'B'} & \epsilon_{C'C'} & t_{C'D'} \\
     t_{D'A} &  t_{D'B} &  t_{D'C} & t_{D'D} & t_{D'A'} &  t_{D'B'} &  t_{D'C'} & \epsilon_{D'D'} \\
  \end{tblr}\right) =: \left(\begin{tblr}{colspec = {c|c},} H_{\mathrm{cell}}^{\mathrm{intra}} & T^{\mathrm{inter}}_{\mathrm{cell}} \\
  \hline
   (T^{\mathrm{inter}}_{\mathrm{cell}})^{\dag} & H_{\mathrm{cell}}^{\mathrm{intra}} \end{tblr}\right)
\end{equation}

Note that this matrix considers only intercell hopping between two adjacent unit cells as well as intracell hopping inside the two unit cells. The hopping rates $t_{ij}$, which are finite for our system, are colour coded (equal colors correspond to equal transition amplitudes in the symmetric case).
In the second equality we have split up the Hamiltonian into intercell and intracell operators.
Given the shape of the unit cell, we can write the Hamiltonian as a tensor product:
\begin{equation}
    H =\sum_n \vert n \rangle \langle n \vert \otimes H_{\mathrm{cell}}^{\mathrm{intra}} 
    \quad + \left( \sum_n \vert n \rangle \langle (n \bmod N)+1 \vert \otimes T^{\mathrm{inter}}_{\mathrm{cell}} + h.c. \right)
\end{equation}

\begin{figure}[h]
  \centering
    \includegraphics[width=0.4\columnwidth]{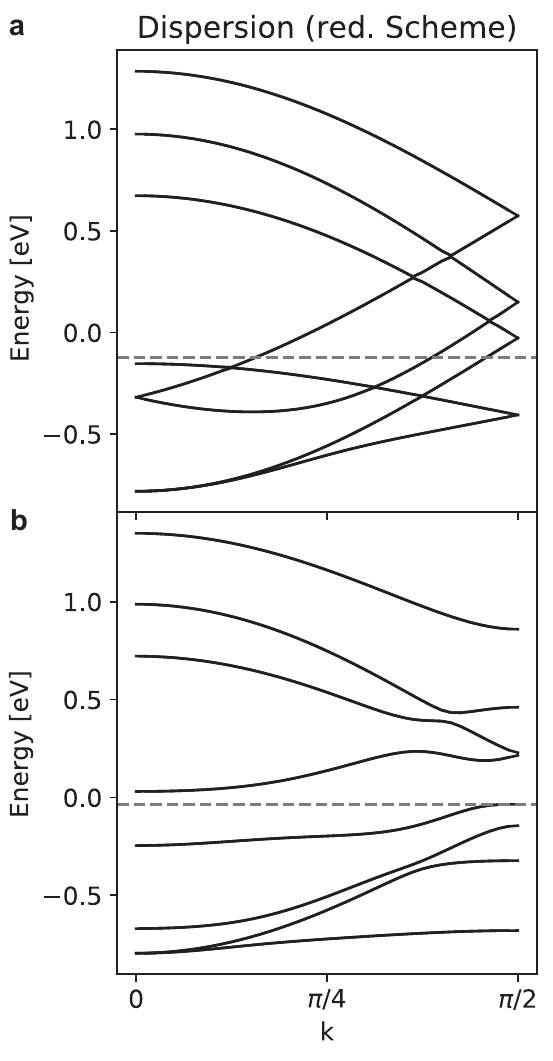}
    \caption{\textbf{Electronic band structure for }(\textbf{a}) the metallic symmetric phase and (\textbf{b}) the distorted insulating phase. Bands are folded back around the $X$-point of the distorted configuration. The respective Fermi energies are shown as a dashed line in both plots.}
\label{fig:dispersion_combined}
\end{figure}
The $n$ mod $N$ term implements the periodic boundary conditions in the direction parallel to the wires (in perpendicular direction, we apply open boundary conditions). 
We use the discrete translational symmetry of the system to calculate the momentum eigenstates using $\vert n \rangle = \frac{1}{\sqrt{N}} \sum_k e^{-ikn} \vert k \rangle$ with $k=\frac{2\pi m}{aN}$ and $m=1,...,N$ and set the distance between two unit cells $a=1$. Note that $k$ here corresponds to $k_{\parallel}$ in Fig. \ref{fig1}c of the main text. Following the naming of transition variables from Ref.~\cite{jeckelmann_grand_2016}, we obtain the Bloch Hamiltonian $H(k)$ for the $(4\times1)$ unit cell:
\begin{equation}\label{eq:H_k}
    H(k) =  \left(\begin{tblr}{colspec = {cccc},}  
    \epsilon_{O} - 2t_{O} \cos{k} & -2t_{IO} e^{\frac{-ik}{2}} \cos{\frac{k}{2}} & 0 & 0\\
   -2t_{IO} e^{\frac{ik}{2}} \cos{\frac{k}{2}} & \epsilon_{I} - 2t_{I2} \cos{k} & -2t_{I1} e^{\frac{ik}{2}} \cos{\frac{k}{2}} & 0 \\
   0 & -2t_{I1} e^{\frac{-ik}{2}} \cos{\frac{k}{2}} & \epsilon_{I} - 2t_{I2} \cos{k} & -2t_{IO} e^{\frac{-ik}{2}} \cos{\frac{k}{2}} \\
   0 & 0 & -2t_{IO} e^{\frac{-ik}{2}} \cos{\frac{k}{2}} & \epsilon_{O} - 2t_{O} \cos{k}\\
   \end{tblr} \right)
\end{equation}

The eigenstates of the entire system are then found via numerical diagonalization of $H(k)$ and calculation of the tensor product with $\vert k \rangle$ for each allowed value of $k$. 
For each energy eigenvector with $H \vert \Psi_n \rangle = E_n \vert \Psi_n \rangle$ we find a corresponding state, constructed from $\vert k \rangle$, and the eigenvectors of $H(k)$, i.e. $\vert \Psi_n \rangle \leftrightarrow \vert k \rangle \otimes \vert i \rangle$ (up to an arbitrary complex phase), where $\vert k \rangle = \frac{1}{\sqrt{N}}\sum_n e^{-ikn}\vert n \rangle $ and $\vert i \rangle $ are the eigenvectors of $H(k)$. 
Due to the dimension of $H(k)$ we have four energies for each possible value of $k$, corresponding to the band structure for the symmetric configuration (Fig. \ref{fig:dispersion_combined}A).
We adapt the parametrization from Ref.~\cite{jeckelmann_grand_2016} to model the metal to insulator phase transition of the indium nanowires.
To this end, two parameters are introduced, corresponding to shear- and rotational structural distortions. The effect on the transition rates is then modeled by:
\begin{equation}
    t_{ij} (d_{ij}) =  t^{\phantom{0}}_{ij} \exp{\left[-\alpha^{\phantom{0}}_{ij} \left( d^{\phantom{0}}_{ij} - d_{ij}^0 \right)\right]} \, ,
    \label{eq:transition}
\end{equation}
where $d_{ij}^0$ and $d_{ij}$ are the distances between lattice sites $i$ and $j$ before and after distortion, respectively. The distortion leads to a unit-cell doubling due to discrete translational symmetry breaking. Fig. \ref{fig:dispersion_combined}b shows the band structure of the symmetry-broken configuration in a reduced scheme, i.e. the band structure is folded back around the $X$-point.

\subsubsection*{Model for a Peierls heterostructure}
\label{sec:model_hetero}
\begin{figure}[!h]
  \centering
    \includegraphics[width=0.9\columnwidth]{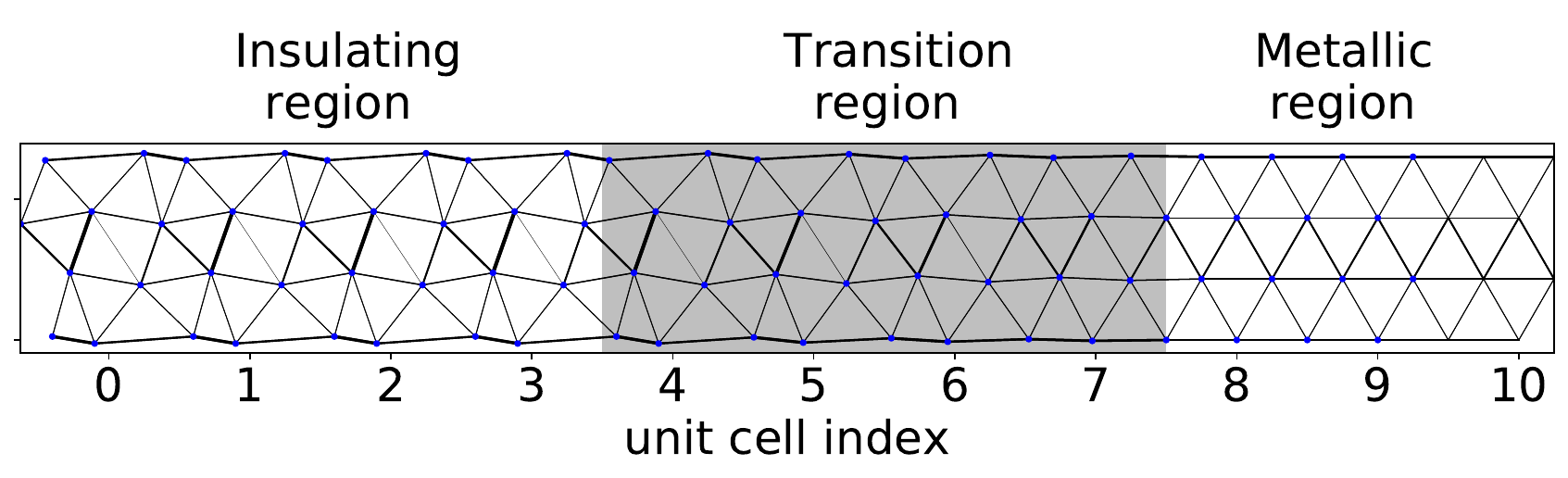}
\caption{\textbf{Lattice for a heterostructure with a transition region extending over 4 unit cells.}}
\label{fig:lattice_hetero}
\end{figure}

The Hamiltonian for the Peierls heterostructure combines the insulating Peierls-distorted phase and the metallic symmetric phase. We model the Hamiltonian as being made up of $(8\times2)$ unit cells where the distortion parameters for each cell can be individually altered and the transition rates change according to Eq.~\ref{eq:transition}. A metallic domain of 20 cells width $\approx$$\,$15.4$\,\mathrm{nm}$ \cite{jeckelmann_grand_2016} is sandwiched between two equally sized homogeneously distorted insulating regions (each of 96 cell width $\approx\,$73.7$\,\mathrm{nm}$). Given the periodic boundary conditions in our model, the two insulating regions should be considered as one region connecting to the metallic domain on both sides. As a side note, the cells in the metallic region contain two symmetric unit cells, i.e. 40 $(4\times1)$ unit cells. Large jumps in the distortion parameters on the phase junction/boundary are avoided by designating a transition region of 4 cells ($\approx\,$3.1$\,\mathrm{nm}$) where the parameters are linearly interpolated (Fig.~\ref{fig:lattice_hetero}). This junction size is derived from experimentally determined values of 0.4-7$\,$nm \cite{song_dynamical_2019}.

The corresponding Hamiltonian is not symmetric under arbitrary lattice translations, hence defining crystal momentum in this system is non-trivial. Instead of investigating the band structure, we calculate the local density of states (LDOS) which we define at cell index $i$ and energy $E$ as $\mathrm{LDOS}(i,E) := \sum_{n, x \,\in \,\mathrm{unit cell}(i)} \left\vert \Psi_{n}(x) \right\vert^2 \delta(E-E_n)$. Since we are grouping 8 sites of a unit cell into one cell index, using the cell index to be the position is only valid as long as the distortion of the unit cells is small. This is the case for our choice of parameters. We broaden the resulting LDOS values with a bivariate Lorentzian with a mean absolute deviation of $\gamma = 0.015$, corresponding to a spatial resolution of approximately 1.5$\,$nm and $3\cdot10^{-3}\,$eV,
\begin{equation}
   \mathrm{LDOS}_{\gamma}(i,E) = \sum_{l,n}  \frac{\gamma \cdot \mathrm{LDOS}(l,E_n)}{2\pi \left(\left(\frac{i-l}{N}\right)^2 + \left(\frac{E-E_n}{E_{\mathrm{max}}-E_{\mathrm{min}}}\right)^2 + \gamma^2 \right)^{3/2}} \, .
\end{equation}

The calculated LDOS for the Peierls heterostructure is depcted in Extended data Fig.~\ref{LDOS}. 

\subsection*{Conductivity in a percolation network}\label{Met_perc}

\begin{figure}[h]%
\centering
\includegraphics[width=0.6\textwidth]{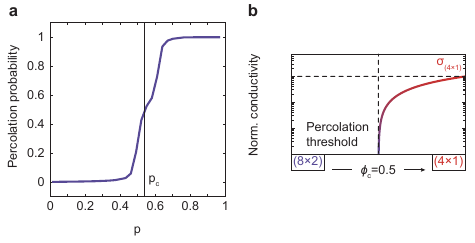}
\caption{\textbf{Voronoi tiling percolation and conductivity in percolation network.} (\textbf{a}), The size of the largest connected component in a randomized Voronoi network is calculated as function of switching probability $p$. (\textbf{b}), Conductivity of the resulting percolative metallic network above the tiling percolation threshold, following a universal scaling law ($\propto(\langle \phi\rangle-\phi_c)^{1.3}$) \cite{sarikhani_unified_2022}.}
\label{fig:Meth_perc}
\end{figure}

The rotational domain texture of indium nanowires \cite{shim_true_2023} is approximated by a randomized Voronoi tiling. To this end, a Voronoi network and corresponding Delauney triangulation is constructed from a Poisson-distributed point grid of $600\times600$ sites. For finding the percolation threshold, we follow established procedures \cite{bollobas_critical_2006, becker_percolation_2009}. An active seed site is placed on the network center and its neighbors are turned on with a probability $p$ or turned off with probability $1-p$ \cite{leath_cluster_1976}. Depending on $p$, the seed will either die off or grow indefinitely, thus marking the transition towards a fully percolated network. Statistical averaging was performed across 500 runs for each $p$ to determine the critical threshold $p_c$ for growth beyond the grid bounds (Fig. \ref{fig:Meth_perc}a). We find $p_c\approx0.54$, in good agreement with the literature value for Voronoi tiling percolation ($p_c=0.5$) \cite{bollobas_critical_2006, becker_percolation_2009}. 

Above the percolation threshold, $p$ approaches the network filling factor $\phi$, i.e. the metallic/insulating surface fraction. The effective conductivity of the network, in this regime, follows a universal scaling law \cite{sarikhani_unified_2022}: 
\begin{equation}
    \sigma_{eff}\propto(\langle \phi\rangle-\phi_c)^{t}
\end{equation}
where $t$ is a critical exponent with universal value of 1.3 for 2D systems (Fig. \ref{fig:Meth_perc}b) \cite{stauffer_introduction_2018}. For the presented simulations, we chose $\langle \phi\rangle=0.6$ , above the percolation threshold.

\subsection*{Electrical anisotropy}\label{Met_anis}

\begin{figure}
\centering
\includegraphics[width=0.6\textwidth]{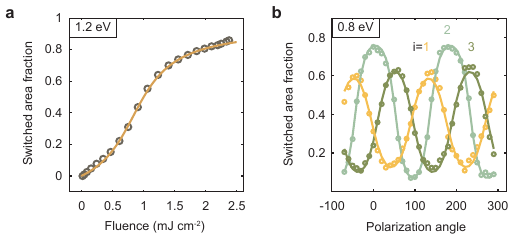}
\caption{\textbf{Fits to experimental data.} (\textbf{a}), Fluence-dependent isotropically switched surface fraction at 1.2\,eV incident photon energy (markers) and fit to the experimental data (line). (\textbf{b}) Polarization-dependent and domain-resolved switched surface fraction at 0.8\,eV incident photon energy (markers) and fit to the experimental data (lines).}
\label{fig:Meth_cond}
\end{figure}

\begin{figure}[h]%
\centering
\includegraphics[width=0.5\textwidth]{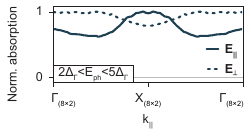}
\caption{\textbf{Polarized optical absorption at near-IR photon energies.} Calculated normalized optical absorption at near-IR photon energies. The absorption is delocalized throughout the electronic band structure and unspecific towards incident polarization.}
\label{abs_highE}
\end{figure}

\begin{figure}[h]%
\centering
\includegraphics[width=0.55\textwidth]{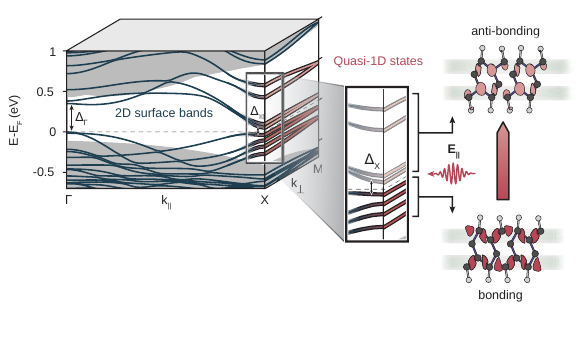}
\caption{\textbf{Band structure and valley-selective photoexcitation.} Left, density-functional-theory (DFT) calculated electronic band structure of the Si(111)(8$\times$2)-In phase (see Methods for details). Gray, projected Si bulk bands; dark green, surface localized bands; red, quasi-1D states with bonding and anti-bonding orbital character (right). $\Delta_{\Gamma}$ and $\Delta_{X}$ denote CDW gaps, leading to the formation of energetic valleys in the band structure. $k_{\vert\vert}$ and $k_{\perp}$ are the momentum vectors along Brillouin zone high-symmetry points with respect to the wire direction.}
\label{band_structure}
\end{figure}

\begin{figure}[h]%
\centering
\includegraphics[width=1\textwidth]{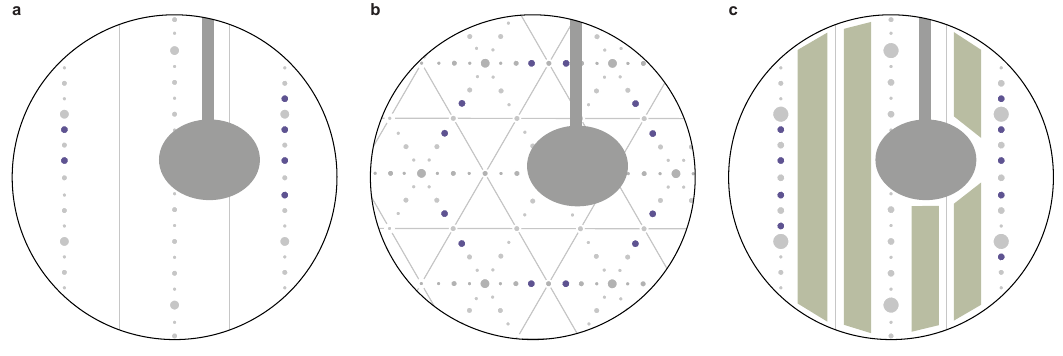}
\caption{\textbf{Analyzed diffraction features.} Schematic diffraction image of the (8$\times$2) indium phase on the (111) face of a Si wafer. (8$\times$2) diffraction spots, analyzed in (\textbf{a}), Fig. \ref{fig2}b on a stepped wafer and \textbf{b}, Fig. \ref{fig2}d on a flat wafer are highlighted (blue). \textbf{c}, (8$\times$2) (blue) and diffuse background (green) features, analyzed in Fig. \ref{fig3}.}
\label{Ext_DiffrFeat}
\end{figure}

\begin{figure}[h]%
\centering
\includegraphics[width=0.55\textwidth]{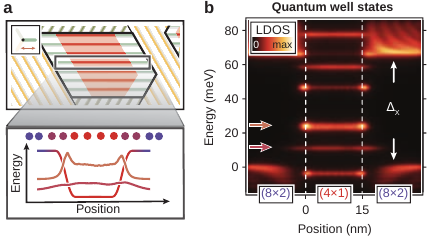}
\caption{\textbf{Tight-binding calculation of Peierls heterostructure local density of states.} (\textbf{a}), Sketch of transient quantum well formation due to suppressed relaxation dynamics and domain-specific switching. Long-lived metallic segments are embedded in relaxed insulating nanowires. (\textbf{b}), Tight-binding calculations of the local density of states along the atomic chain show the formation of quantized metallic states within the insulating band gap ($\Delta_X$). In the transition region one expects strong band bending and therefore high curvature, which increases the effective mass of the electron and causes localization. Mismatch in band alignment between the metallic and insulating phase further leads to confinement even beyond the insulating gap in metallic and insulating segments.}
\label{LDOS}
\end{figure}

The electrical properties of the percolated network are modeled as an ensemble average of metallic rotational domains, determined from experimental data. To this end, we consider the polarization dependence of switched nanowires (see also Fig. \ref{fig2}d in the main text) to depend both on the relative absorption as well as on the non-linear switching yield with absorbed fluence. Fig. \ref{fig:Meth_cond}a shows the fluence-dependent switched area fraction $\phi_i(F)$ for 1.2$\,$eV photon energy, where $i$ denotes the nanowire orientation. Even though the following calculations are performed for 0.8$\,$eV, the threshold characteristic was found to be universal in the investigated energy range. A best fit to the data was found for the phenomenological expression:
\begin{equation}
    \phi_{i}(F)=a_{th}\cdot\arctan((F-f_{th})/w_{th})+c_{th}
\end{equation}
where $a_{th}$, $f_{th}$, $w_{th}$ and $c_{th}$ denote fit parameters for the threshold amplitude, fluence, width and central value and $F$ is the incident fluence. For 1.2$\,$eV, the nanowire absorption anisotropy is completely diminished by charge carrier delocalization across all domains. For 0.8$\,$eV, however, the absorbed fluence and switched area fraction become orientation-specific. Hence, the polarization dependence of the switched area fraction is fit by an anisotropic absorption:
\begin{equation}
    \phi_{i}(F,\varphi)=a_{th}\cdot\arctan((F\cdot(A_{||}cos(\varphi-\varphi_0)^2+A_{\perp}sin(\varphi-\varphi_0)^2)-f_{th})/w_{th})+c_{th}
\end{equation}
where $A_{||}$, $A_{\perp}$ and $\varphi_0$ are fit parameters for the relative absorption and nanowire orientation with respect to the in-plane electric field and $\varphi$ denotes the incident polarization angle. Fits to the experimental data are depicted in Fig. \ref{fig:Meth_cond}b.

In the simulation of a percolative conduction network, we choose a corresponding $F$ for every $\varphi$ so that the mean switched surface area:
\begin{equation}
    \langle \phi\rangle=\sum_{i=1}^3 \phi_{i}(F_{0.6},\varphi)/3
\end{equation}
reaches a constant value of 0.6, well above the percolation threshold (see Methods: Conductivity in a percolation network). In this way, the composition of the metallic network of rotational domains is weighted by the orientation-resolved relative fraction of switched domains ($\phi_{i}(F_{0.6},\varphi)/(3\langle\phi\rangle)$). \\
Based on the experimental data, we calculate the anisotropic resistivity tensor of the rotational domain texture from the ensemble mean. To this end, we follow established procedures for anisotropic conductive fillers, embedded in an insulating matrix \cite{kumar_evaluating_2016, tarasevich_effective_2023}.\\
The intrinsic in-plane resistivity tensor of indium nanowires is:
\begin{equation}
\rho=\left( \matrix{ \rho_{||} & 0 \cr
0 & \rho_\perp \cr} \right)
\end{equation}
where $\rho_{||}=1/\sigma_{||}=1.4\times10^{3}\,\Omega/square$ and $\rho_{\perp}=1/\sigma_{\perp}=8.3\times10^{4}\,\Omega/square$ are taken from literature \cite{kanagawa_anisotropy_2003}. Within the ensemble of rotational domains, the mean resistivity $\langle\rho\rangle$ is expressed in terms of a tensor rotation by the angle of nanowires $\theta$ with respect to the above parallel axis \cite{frenkel_evidence_1985, wu_angle-resolved_2020}:
\begin{equation}
    \bar{\rho}=(\rho_{||}+\rho_\perp)/2
\end{equation}
\begin{equation}
    \Delta\rho=(\rho_{||}-\rho_\perp)/2
\end{equation}
\begin{equation}
\langle\rho\rangle=\left( \matrix{ \bar{\rho}+\Delta\rho\langle \cos(2\theta)\rangle & \Delta\rho\langle \sin(2\theta)\rangle \cr
\Delta\rho\langle \sin(2\theta)\rangle & \bar{\rho}-\Delta\rho\langle \cos(2\theta)\rangle} \right)
\end{equation}
The angular dependence is calculated with respect to one nanowire orientation ($i=2$ in fig.  \ref{fig:Meth_cond}B), such that $\theta_1=-2\pi/3$, $\theta_2=0$ and $\theta_3=2\pi/3$:
\begin{equation}
    \langle \cos(2\theta)\rangle=(\phi_{1}(F_{0.6},\varphi)\cos(2\theta_1)+\phi_{2}(F_{0.6},\varphi)\cos(2\theta_2)+\phi_{3}(F_{0.6},\varphi)\cos(2\theta_3))/3\langle\phi\rangle
\end{equation}
\begin{equation}
    \langle \sin(2\theta)\rangle=(\phi_{1}(F_{0.6},\varphi)\sin(2\theta_1)+\phi_{2}(F_{0.6},\varphi)\sin(2\theta_2)+\phi_{3}(F_{0.6},\varphi)\sin(2\theta_3))/3\langle\phi\rangle
\end{equation}
Diagonalization of the mean resistivity yields the anisotropy ratio, i.e. the ratio of diagonal components, with respect to the polarization-dependent principal axes (Fig. \ref{fig4}c in the main text). In the depicted Hall bar geometry, the current $\textbf{J}$ is constrained in the x-direction ($\textbf{J}=J_x$). The resulting electric field ratio is thus calculated as:
\begin{equation}
    E_{y}/E_{x}=\langle\rho\rangle_{xy}/\langle\rho\rangle_{xx}=\Delta\rho\langle \sin(2\theta)\rangle/(\bar{\rho}+\Delta\rho\langle \cos(2\theta)\rangle)
\end{equation}

\newpage

\bibliography{bibliography}

\bibliographystyle{sn-standardnature}

\newpage

\section*{Extended data figures}\label{Ext}

\newpage
\newpage

\newpage


\section*{Acknowledgments}
This work was funded by the European Research Council (ERC Advanced Grant “ULEEM,” ID: 101055435) and 
by the Deutsche Forschungsgemeinschaft (DFG, German Research Foundation) - 217133147/SFB 1073, projects A05 and B03.

\newpage

\section*{Supplementary information}

\subsection*{Simulation results for TB model}
In the presented tight-binding calculations, the complexity arising from time-dependent optical intensity or the manifold of structural degrees of freedom acting in concert, is omitted. However, a qualitative understanding of valley-selective optical absorption, arising from the formation of a CDW gap is captured by Eq.~\ref{eq:ham_position} in the Methods section.

\subsubsection*{Current operator}
In order to discuss the currents and the associated absorption anisotropies (see Methods section \ref{tb_sim}), we treat the system in the two-dimensional plane with coordinates $(x,y)$ and unit vectors $\hat{x}$ and $\hat{y}$ for both directions, respectively.
We can then define currents in the system, which are parallel to the chain directions (longitudinal direction, denoted by $\hat{x}$) and orthogonal to it (transverse direction, denoted by $\hat{y}$). 
Following Ref.~\cite{yamamoto_optical_2008}, we can formulate a current operator for each direction:
\begin{equation}
    J_x = \frac{N_{\perp}}{Na}  \sum_{ij} t_{ij}(d_{ij}) \, \left(\hat{t}_{ij} \cdot \hat{x} \right) \left( c^{\dagger}_{i} c^{\phantom \dagger}_{j} - c^{\dagger}_{j} c^{\phantom \dagger}_{i} \right)  \, ,
\end{equation}
where $\hat{t}_{ij}$ is the unit vector pointing in the direction of the bond connecting lattice sites $i$ and $j$ and $N_{\perp}$ is the number of chains perpendicular to the chain direction, i.e. four in this model.
Analogously, for the orthogonal direction we replace $\hat{x}$ by $\hat{y}$. We use the expression for the optical conductivity derived in Ref.~\cite{gebhard_optical_1997-1} and adapt it to account for half-filling. In this way, we calculate the absorption spectrum as the real part of the optical conductivity $\sigma(\omega)$: 
\begin{equation}
     \sigma_{x/y}(\omega) =  \frac{\pi}{\omega} \, \mathrm{Im} \left[ \! \! \sum_{\fontsize{8}{0}\selectfont\begin{array}{c} n_0\leq \frac{N}{2} \\ n_{\mathrm{ex}}>\frac{N}{2} \end{array}} \! \! \! \!
       \left\vert \langle \Psi_{n_0} \vert \hat{J}_{x/y} \vert \Psi_{n_{\mathrm{ex}}} \rangle \right\vert^2 \! \left( \frac{1}{\omega + (E_{n_{\mathrm{ex}}}\!-E_0)+i\gamma}  
       - \frac{1}{\omega - (E_{n_{\mathrm{ex}}}\!-E_0)+i\gamma} \right) \! \right]
\end{equation}

\begin{figure}[h!]
    \centering
    \includegraphics[width=0.75\columnwidth]{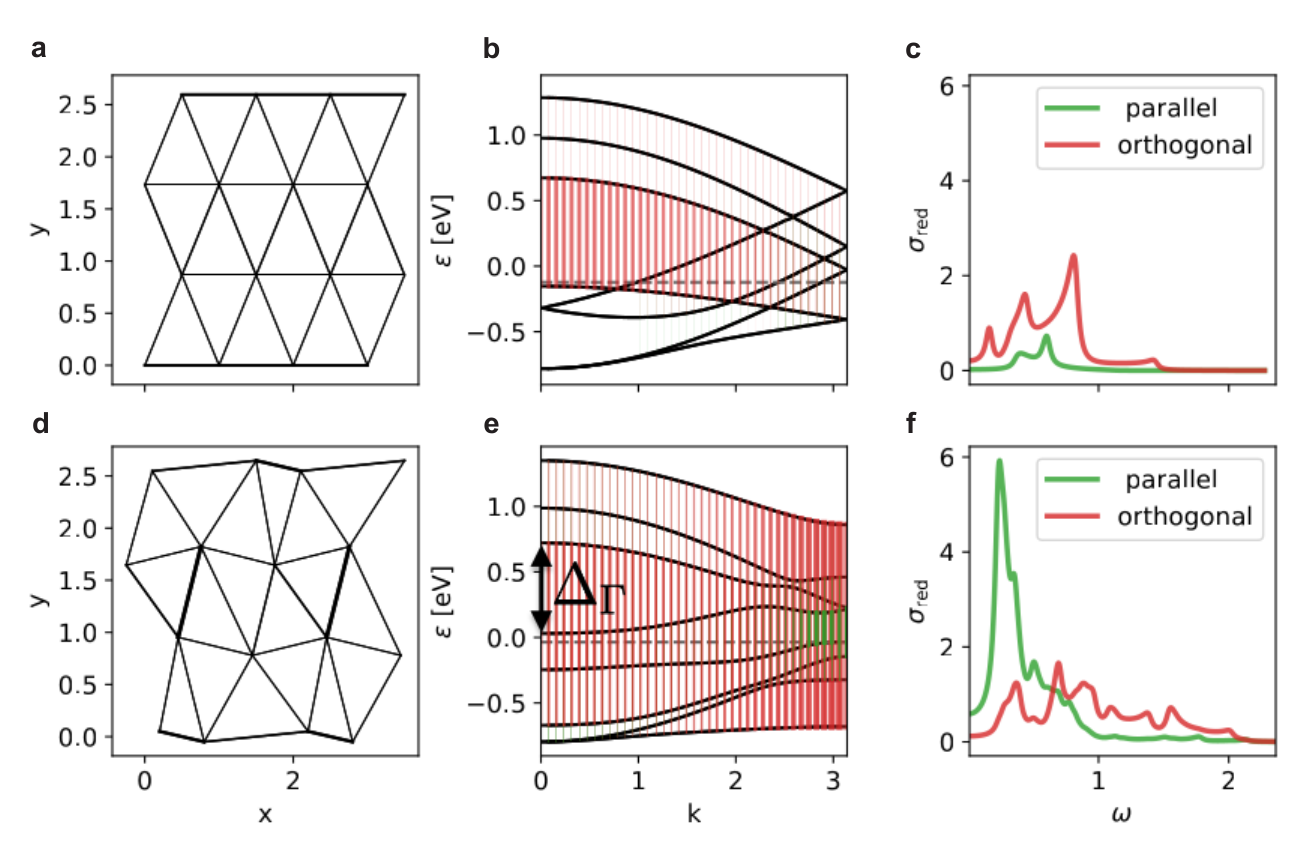}
    \caption{\textbf{Tight-binding simulation results for homogeneous systems.} (\textbf{a,d}) Lattice configuration in real space, (\textbf{b,e}) band structure with weighted transitions and (\textbf{c,f}) optical conductivities for the metallic phase and insulating phase, respectively. $\Delta_{\Gamma}$ corresponds to the energy gap in Fig.~\ref{fig1}c}
    \label{fig:sim_homog_tiles}
\end{figure}

\subsubsection*{Optical conductivities}

We calculate the optical conductivities from optical transitions in the band structure for both the symmetric and distorted nanowire structures  (Fig.~\ref{fig:sim_homog_tiles}). The current operator matrix elements $|\langle \Psi_{n_0} \vert \hat{J}_{x/y} \vert \Psi_{n_{\mathrm{ex}}} \rangle |^2$ of individual transitions are depicted in Fig.~\ref{fig:sim_homog_tiles}b and e, while optical conductivities for both phases are shown in Fig.~\ref{fig:sim_homog_tiles}c and f. In qualitative agreement with DFT calculations (see Fig.~\ref{fig1}c of the main text), we find a strongly anisotropic absorption for the insulating phase at the zone-boundary CDW gap along the nanowire direction.

\subsection*{Coherent control at IR- and near-IR excitation}\label{SI_CC}

\begin{figure}
\centering
\includegraphics[width=0.6\textwidth]{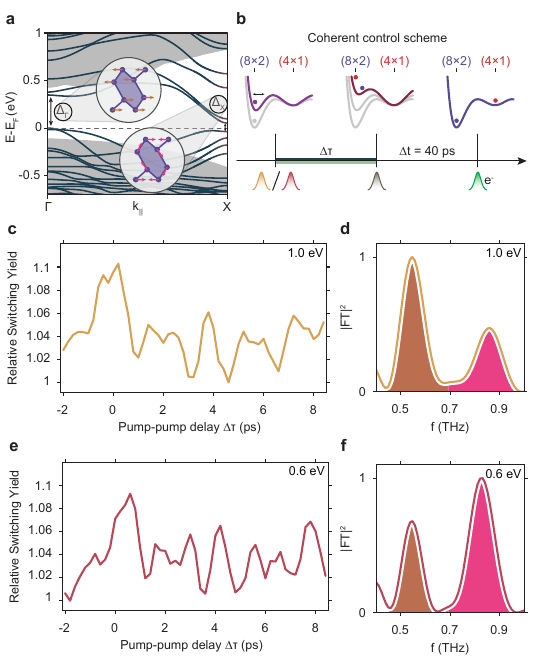}
\caption{\textbf{Photon-energy-resolved vibrational coherences in coherent control measurements.} (\textbf{a}), Calculated band structure along the momentum direction, parallel to the atomic wire. CDW gaps at the $\Gamma$- and $X$-points preferentially couple the shear and rotation vibrational amplitude modes, respectively. (\textbf{b}), Scheme for coherent control measurements. A first weaker preparation pulse creates a coherent vibrational oscillation in the ground state (left), which is translated into a measurable difference in ensemble averaged switching efficiency, depending on the momentary vibrational state upon the arrival of a stronger switch pulse ($E_{ph}$=1.2$\,$eV) after time $\Delta\tau$ (center). The resulting switched surface fraction is probed by a later electron pulse at $\Delta$t=40$\,$ps (right). (\textbf{c}), Relative switching yield as function of $\Delta\tau$ for a preparation pulse photon energy of 1.0$\,$eV, showing characteristic vibrational oscillations. (\textbf{d}), The corresponding Fourier transform reveals the relative amplitudes at the shear mode (0.55 THz) and rotation mode frequencies (0.82 THz) \cite{wippermann_entropy_2010,speiser_surface_2016}. (\textbf{e}), Analogous measurement for a preparation pulse photon energy of 0.6$\,$eV (\textbf{f}), The Fourier transform of the time trace shown in (\textbf{e}).}
\label{SI_fig1}
\end{figure}

A direct fingerprint of the involved electronic states is given by their preferential coupling to the individual amplitude phonon modes that modulate the structural distortion upon excitation \cite{wippermann_entropy_2010,jeckelmann_grand_2016,bockmann_mode-selective_2022}. Specifically, DFT calculations show that the electronic rearrangement due to the population of states at the $\Gamma$-point initiates a shearing between coupled indium chains while the depopulation of states around the X-point drives a hexagon rotary motion (Fig. \ref{SI_fig1}a). We detect the resulting structural dynamics from an optical excitation in a coherent control scheme \cite{horstmann_coherent_2020}, where a weak first pulse of variable wavelength creates vibrational coherence within the (8$\times$2) minimum of the potential energy landscape along the structural distortion (Fig. \ref{SI_fig1}b). Subsequently, a stronger near-infrared pulse transforms a fraction of indium atomic wires into the (4$\times$1) phase, as probed by an electron pulse at $\Delta$t=40$\,$ps. The switched surface fraction thus becomes dependent on the momentary vibrational state and its impact on overcoming the potential energy barrier \cite{bockmann_mode-selective_2022}. Oscillations in the relative switching yield with delay time between optical pulses reveal the vibrational modes, excited by 1.0$\,$eV photons (Fig. \ref{SI_fig1}c). A Fourier transform shows contributions from both shear and rotary modes, as previously observed for higher photon energies \cite{horstmann_coherent_2020}, suggesting relevant transitions within the entire surface Brillouin zone (Fig. \ref{SI_fig1}d) \cite{nicholson_beyond_2018}. The relative strength of the shear mode is likely attributed to the stronger electron-phonon coupling of zone-center electronic states \cite{jeckelmann_grand_2016}. On the other hand, spectral transitions from 0.6$\,$eV photons are highly localized at the $X$-point, thus coupling primarily to the rotation mode, as manifested in the oscillatory component of the relative switching yield (Fig. \ref{SI_fig1}e) and the corresponding Fourier transform (Fig. \ref{SI_fig1}f). The inferred decisive role of charge carriers around the $X$-point for the phase transition is in perfect agreement with photoemission spectroscopy and molecular dynamics simulations \cite{nicholson_beyond_2018}. 

\end{document}